\documentclass[pdflatex,sn-mathphys-num]{sn-jnl}

\usepackage{graphicx}%
\usepackage{multirow}%
\usepackage{amsmath,amssymb,amsfonts}%
\usepackage{amsthm}%
\usepackage{mathrsfs}%
\usepackage[title]{appendix}%
\usepackage{xcolor}%
\usepackage{textcomp}%
\usepackage{manyfoot}%
\usepackage{booktabs}%
\usepackage{algorithm}%
\usepackage{algorithmicx}%
\usepackage{algpseudocode}%
\usepackage{listings}%
\usepackage{bm}%
\usepackage{makecell}%
\usepackage{siunitx}%
\usepackage{array}%
\usepackage{tabularx}%

\theoremstyle{thmstyleone}%

\theoremstyle{thmstyletwo}%

\theoremstyle{thmstylethree}%

\begin{document}

\title[Dual-moon forced dynamics and nonlinear aggregation in Saturn's F ring: From quasi-periodicity to modulated oscillations]{Dual-moon forced dynamics and nonlinear aggregation in Saturn's F ring: From quasi-periodicity to modulated oscillations}

\author*[1]{\fnm{Omar El} \sur{Deeb}}\email{omar.el-deeb@warwick.ac.uk}

\affil*[1]{\orgdiv{Department of Mathematics}, \orgname{University of Warwick}, \orgaddress{\city{Coventry}, \country{UK}}}

\abstract{We develop a minimal nonlinear model to investigate the oscillatory dynamics of Saturn's F ring under dual-moon forcing from Prometheus and Pandora. The model extends classical predator--prey dynamics by incorporating both a nonlinear mass aggregation term $kM^n$ and explicit dual-frequency forcing, capturing how higher-order coagulation physics interacts with multi-moon perturbations. Through systematic numerical integration and dynamical-systems tools, including time-series, spectral, stroboscopic, and rotation number analysis, we identify distinct dynamical regimes controlled by the parameters $n$ and $k$. 

For moderate nonlinearity \((n=1.28,k=0.54)\), the numerical diagnostics are
consistent with bounded quasiperiodic motion, characterized by smooth amplitude
modulation, thin stroboscopic loops, and discrete spectral peaks. For stronger
nonlinearity \((n=1.30,k=0.62)\), the same diagnostics indicate a transition
toward strongly modulated oscillations, with broadened stroboscopic bands and
sideband-rich spectra. A projected rotation-number diagnostic reveals organized
regions in parameter space, including smooth quasiperiodic-like domains and
near-locking bands analogous to Arnold tongues.

Our results show that a reduced deterministic dual-forcing model can generate bounded quasiperiodic-like and strongly modulated aggregation-fragmentation cycles with timescales comparable to relevant moon-ring forcing periods,  providing a possible low-dimensional mechanism contributing to F-ring variability.}

\maketitle

%=============================================================
\section{Introduction}
Saturn, the second-largest planet in our solar system, concentrates a remarkable diversity of physical processes in a single system. Its low density, rapid rotation, deep hydrogen--helium atmosphere, powerful magnetosphere, and spectacular rings make it a natural laboratory for planetary physics and celestial mechanics \cite{colwell_structure_2009, cook_saturns_1973}. The system hosts dozens of moons spanning geologic extremes, from Titan's dense atmosphere to Enceladus's plumes. Observations across wavelengths reveal active weather, seasonal variability, and ring--moon coupling that continually reshapes the environment \cite{blanc_understanding_2025}. Studying Saturn refines theories of disk dynamics, accretion, and resonance which are key ideas for understanding planetary systems and disks.

The dynamics of Saturn's rings arise from the interplay of self-gravity, collisions, viscous diffusion, and time-dependent perturbations from embedded and nearby moons. Particles span micron dust to multi-meter aggregates, with optical depth and size distributions varying radially and azimuthally. In this granular disk, inelastic collisions dissipate energy while collective self-gravity promotes transient clumping. Nonlinear processes such as shear reversal and swing amplification can magnify small perturbations \cite{cuzzi_evolving_2010, cuzzi_saturns_2014, tajeddine_what_2017}. Stochastic impacts and meteoroid bombardment inject fine dust, modifying scattering and collisional regimes \cite{rein_stochastic_2010}. Overlapping resonances and variable driving from multiple moons can create interference patterns and beats, yielding envelopes in clump amplitude and phase lags between mass buildup and velocity dispersion. This complexity motivates reduced models that capture essential feedback while remaining tractable.

Located near the Roche limit \cite{williams_roche_2003}, the F ring balances tidal disruption against self-gravity and collisional sticking, making it an ideal setting to study aggregation-fragmentation cycles \cite{brilliantov_steady_2018}. Encounters with Prometheus are associated with channels, streamers and dense knots, while Pandora provides an additional weaker perturbation to the outer edge, producing quasi-periodic patterns \cite{poulet_dynamical_2001, farmer_understanding_2006}. The ring highlights the sensitivity of granular disks to resonant, time-dependent forcing.

The Cassini--Huygens mission transformed our understanding of Saturn's system through continuous observation \cite{ingersoll_cassini_2020, spilker_cassini-huygens_2019}. Cassini's instruments mapped ring structure at high resolution, tracked temporal variability, and linked features to moon encounters. Imaging revealed multiple F-ring strands, channel formation by Prometheus, knotty clumps, and episodic jets, while stellar occultations constrained particle size distributions and optical depths. Spectroscopy distinguished water-ice signatures and traced dust production. Radar and radio science probed ring mass and microstructure, and gravity measurements refined Saturn's interior and the moons' masses. Long-baseline monitoring resolved seasonal effects and synodic recurrences, enabling quantitative links between resonance locations, driving periodicities, and observed morphologies across the ring system.

Early theories of ring origin emphasized catastrophic or collisional scenarios, proposing that rings form from the disruption of a progenitor body or the incomplete accretion of satellite material within the Roche zone. Lumme \cite{lumme_formation_1972} developed an influential framework linking ring formation to tidal fragmentation and collisional evolution, highlighting how Saturn's gravity can inhibit accretion and promote sustained particle populations. Decades later, Canup \cite{canup_origin_2010, murray_origin_2018} advanced a comprehensive model in which a Titan-sized satellite migrated inward and experienced mass removal through tidal stripping near Saturn's Roche limit, naturally generating today's massive inner rings and potentially seeding the small inner moons.

Despite extensive imaging and theoretical work, there remains a need for reduced dynamical models that isolate how nonlinear aggregation and explicit dual-moon forcing interact in the F ring. In particular, a compact, tractable two-variable model combining a tunable nonlinear aggregation law with explicit Prometheus-Pandora forcing can complement existing spatial and stochastic descriptions \cite{alrebdi2025saturn, esposito2025statistics}. Prior predator--prey models \cite{diz-pita_predatorprey_2021} successfully reproduced clumping driven by a single periodic perturber, but typically linearized mass growth or neglected multi-frequency effects \cite{esposito_predatorprey_2012}. Moreover, several physical mechanisms like gravitational compression, shear reversal, and self-gravity wakes \cite{ghigliotti_shear-reversal_2015, nicholson_self-gravity_2010, colwell_self-gravity_2007, tiscareno_analytic_2010} can induce nonlinear mass-dependent growth, which a linear term cannot represent. Filling this gap allows us to assess regimes that sustain bounded quasiperiodic-like modulation, predict beat-like envelopes, and connect measurable timescales with the underlying microphysics.

We build on the prior clumping--dispersal framework that tracks two coupled
variables, the mean aggregate mass and the squared relative velocity dispersion \cite{esposito_predatorprey_2012, el_deeb_higher_2024}. Our contribution is a parsimonious extension that preserves these minimal state variables and the core aggregation-fragmentation feedback while introducing two explicit drivers: Prometheus, anchored to the observed synodic cadence used in earlier F-ring modelling, and Pandora, represented as a weaker secondary modulation estimated from mean motions. This addition lets us examine how multifrequency forcing \cite{nayfeh2024nonlinear, mitsi1998dynamics} organizes the dynamics without inflating dimensionality or obscuring mechanisms. The goal is not to provide a direct fit to Cassini images or occultation profiles, but to determine whether this reduced deterministic system can generate bounded modulation, quasiperiodic-like response, and near-locking behavior on physically relevant forcing timescales. The resulting outputs, such as modulation periods, spectral sidebands, phase lags, and admissible boundedness regions, define quantities that can be compared with Cassini-derived aggregate or brightness proxies in future work.

The methods of dynamical systems provide a powerful tool that has numerous applications in mathematical and quantitative models \cite{huang_generic_2009, gotoda_detection_2014, dibeh_synchronization_2024, massar_equilibrium_2025, hamzi_learning_2021, li_data-driven_2021}. In our context, they constitute a principled way to reduce the complex physics of rings into interpretable low-dimensional models with predictive structure. When multiple frequencies act, reduced nonlinear systems may display beat phenomena, quasiperiodic-like response, near-locking, and transitions between different modulation regimes. Nonlinear parameterizations such as a power-law growth term allow us to embed microphysical effects within a tractable mean-field description and to map parameter regimes in which bounded long-term oscillations persist, and where the physically admissible domain remains.

The paper is structured as follows: In Section II, we present the dual-moon forced model. Section III details the resulting dynamics using time-series, spectral, and stroboscopic diagnostic tools. We also discuss the physical implications of our findings and the limitations of our work in the same section before presenting our conclusions in Section IV.

%===========================================================
\section{The Model}
The model builds on the clumping-dispersal framework \cite{esposito_predatorprey_2012, el_deeb_higher_2024} for the clumping dynamics of Saturn's F ring and extends it to examine the response of the aggregation-fragmentation cycle to two prescribed moon-ring forcing frequencies. Prometheus' periodic encounters with the ring's core are known from Cassini observations to trigger the formation of channels, streamers, and recurrent clumps. Pandora, orbiting just outside the F ring, contributes additional, though weaker perturbations, allowing us to test how an additional nearby forcing frequency influences the long-term oscillatory cycle.

The model describes the coupled evolution of two quantities: the mean aggregate mass $M(t)$ of the ring particles and the square of their relative velocity dispersion $V(t)$. Their temporal behavior arises from coagulation, fragmentation, viscous stirring, and external forcing by the two moons. The nonlinear coagulation term $kM^{n}$ phenomenologically represents enhanced aggregation efficiency from microphysical processes such as gravitational instability, shear reversal, and swing amplification \cite{el_deeb_higher_2024, ross_predator-prey_2016}. The parameters \(k\) and \(n\) are not independently fitted microphysical constants in the present study. They are control parameters used to explore how the reduced aggregation-fragmentation cycle changes as nonlinear mass-dependent growth is strengthened. The fragmentation term $-\,(V/v_{\mathrm{th}}^{2})(M/T)$ accounts for the breakup of aggregates when their relative velocity exceeds the sticking threshold $v_{\mathrm{th}}$. Appendix A provides an explicit derivation of the two baseline equations, the time scaling, and the roles of \(M_0\) and \(v_{\rm esc}\).

Including the two periodic forcing terms gives the coupled nonlinear system
\begin{equation}
\begin{aligned}
\frac{dM}{dt} &= 
\frac{M}{T}
- \frac{V}{v_{\mathrm{th}}^{2}}\frac{M}{T}
+ kM^{n}, \\[6pt]
\frac{dV}{dt} &= 
-\frac{(1-\epsilon^{2})}{T}\,V
+ \frac{M^{2}}{M_{0}^{2}}\frac{v_{\mathrm{esc}}^{2}}{T_{\mathrm{orb}}}
- \Big[A_{0}\cos(\omega_{1}t) + A_{0}\eta\,\cos(\omega_{2}t)\Big],
\end{aligned}
\label{eq:two-moon}
\end{equation}
where the first cosine term represents the Prometheus forcing and the second, scaled by the dimensionless factor $\eta\in[0,1]$, represents the Pandora forcing.
The overall negative sign in the forcing contribution follows the phase
convention discussed in Appendix \ref{app:model-derivation}: the reference phase is chosen so that \(\cos(\omega_i t)>0\) corresponds to the streamline-crowding part of the
encounter, which reduces the relative velocity dispersion and promotes
aggregation. A different choice of phase origin would absorb this sign into a
phase shift. 

All integrations use nondimensional time \(t/T_{\rm orb}\), so \(T_{\rm orb}=1\) in Eq.~(1). The physical orbital period at the F-ring core is used only when converting model periods into days. For radius $a\approx1.4018\times10^{5}\,\mathrm{km}$, we get that:
\begin{equation}
T_{\mathrm{orb}}^{\rm phys} = 2\pi\sqrt{\frac{a^{3}}{\mu_{\mathrm{Saturn}}}}
\approx 0.620~\mathrm{days}.
\label{eq:torb}
\end{equation}

The synodic period for each moon--ring interaction is determined from their angular velocities:
\begin{equation}
T_{s,i} = \frac{2\pi}{\big|\Omega_{\mathrm{ring}} - \Omega_{i}\big|}
= \frac{1}{\left|\frac{1}{T_{\mathrm{ring}}} - \frac{1}{T_{i}}\right|},
\label{eq:ts}
\end{equation}
and the corresponding angular frequencies are $\omega_{i}=2\pi/T_{s,i}$. In numerical experiments,  we adopt
\begin{equation}
T_{s,1}=112\,T_{\mathrm{orb}},\qquad
T_{s,2}=71\,T_{\mathrm{orb}},
\label{eq:tsvalues}
\end{equation}
based on Cassini's observed Prometheus recurrence \cite{esposito_predatorprey_2012} and the synodic estimate for Pandora.

\begin{table}[t]
\caption{Parameters used for numerical solutions of model~(\ref{eq:two-moon}) with two-moon forcing.}
\label{tab:params}
\begin{tabular}{llll}
Parameter & Symbol & Value & Description \\
\hline
Orbital period (F ring)      & $T_{\mathrm{orb}}$ & $1$                      & Reference orbital period \\
Optical depth                & $\tau$             & 0.1                      & Local optical depth \\
Collision period             & $T=T_{\mathrm{orb}}/(4\tau)$ & 2.5          & Mean collision time \\
Coefficient of restitution   & $\epsilon$         & 0.6                      & Collisional dissipation \\
Reference aggregate mass     & $M_{0}$            & $2\times10^{9}$ g        & 10 m ice clump, $\rho=0.5$ g\,cm$^{-3}$ \\
Escape velocity              & $v_{\mathrm{esc}}$ & 0.5 m\,s$^{-1}$          & From surface of $M_{0}$ \\
Prometheus forcing amplitude & $A_{0}$            & 0.1                      & Forcing magnitude \\
Pandora relative amplitude   & $\eta$             & 0.25                     & Forcing amplitude $A_{0}\eta$ \\
Sticking threshold velocity  & $v_{\mathrm{th}}$  & 1.33                     & Aggregation threshold \\
Prometheus forcing period    & $T_{s,1}$          & $112\,T_{\mathrm{orb}}$  & Cassini recurrence \\
Pandora forcing period       & $T_{s,2}$          & $71\,T_{\mathrm{orb}}$   & Synodic estimate \\
\end{tabular}
\end{table}

\paragraph*{Parameters.}
Time is scaled by $T_{\mathrm{orb}}$; the local optical depth $\tau$ fixes the mean collision time $T=T_{\mathrm{orb}}/(4\tau)$, which controls both mass exchange and collisional damping in $\dot M$ and $\dot V$. Dissipation in $\dot V$ scales with the coefficient of restitution $\epsilon$, while viscous stirring is referenced to the escape-velocity scale $v_{\mathrm{esc}}$ for an aggregate of mass $M_{0}$. External driving enters $\dot V$ explicitly as $A_{0}\cos(\omega_{1}t)+A_{0}\eta\cos(\omega_{2}t)$ with synodic periods (\ref{eq:tsvalues}). The sticking threshold $v_{\mathrm{th}}$ sets the velocity beyond which fragmentation dominates aggregation in $\dot M$. Higher-order aggregation is represented by \(kM^n\). The exponent \(n>1\) controls the degree of superlinear mass dependence, while \(k\) controls the effective strength of this phenomenological pathway.

For each parameter pair \((n,k)\), Eq.~(1) is integrated forward from positive
initial conditions and monitored for physical admissibility. A trajectory is
classified as physically admissible only while \(M(t)\geq0\) and \(V(t)\geq0\)
and while both variables remain bounded. Parameter combinations that collapse
to a near-zero mass state, cross \(V<0\), cross $M<0$  or exceed the
imposed boundedness threshold (including diverging solutions) are not interpreted as Saturn-like long-term
ring dynamics. They are instead classified  as having no physically admissible
long-term solution in the reduced model, and labelled "No Physical Solution" in the figures.

\subsection{Numerical implementation}

All numerical integrations were performed in Mathematica using \texttt{NDSolve}
with the \texttt{StiffnessSwitching} method. The baseline computations used
\texttt{WorkingPrecision -> MachinePrecision}, \texttt{AccuracyGoal -> 10},
\texttt{PrecisionGoal -> 10}, \texttt{MaxSteps -> Infinity}, and
\texttt{MaxStepFraction -> 1/50}. Long trajectories were sampled once every
Prometheus synodic period \(T_{s,1}\) after discarding an initial transient.

Representative parameter values were recomputed using higher precision and
smaller maximum step fractions to check that the qualitative regime
classification was not a numerical artifact. The two bounded regimes used in
the main figures remained qualitatively unchanged under these checks, while
parameter values classified as ``No Physical Solution'' continued to fail to
produce physically admissible long-term bounded trajectories. The robustness
outcomes for representative bounded and non-admissible parameter
values are reported in Appendix~B, Table~\ref{tab:numerical-robustness}.

%=============================================================
\section{Results and discussion}

\begin{figure}[t]
    \centering
    \includegraphics[width=\linewidth]{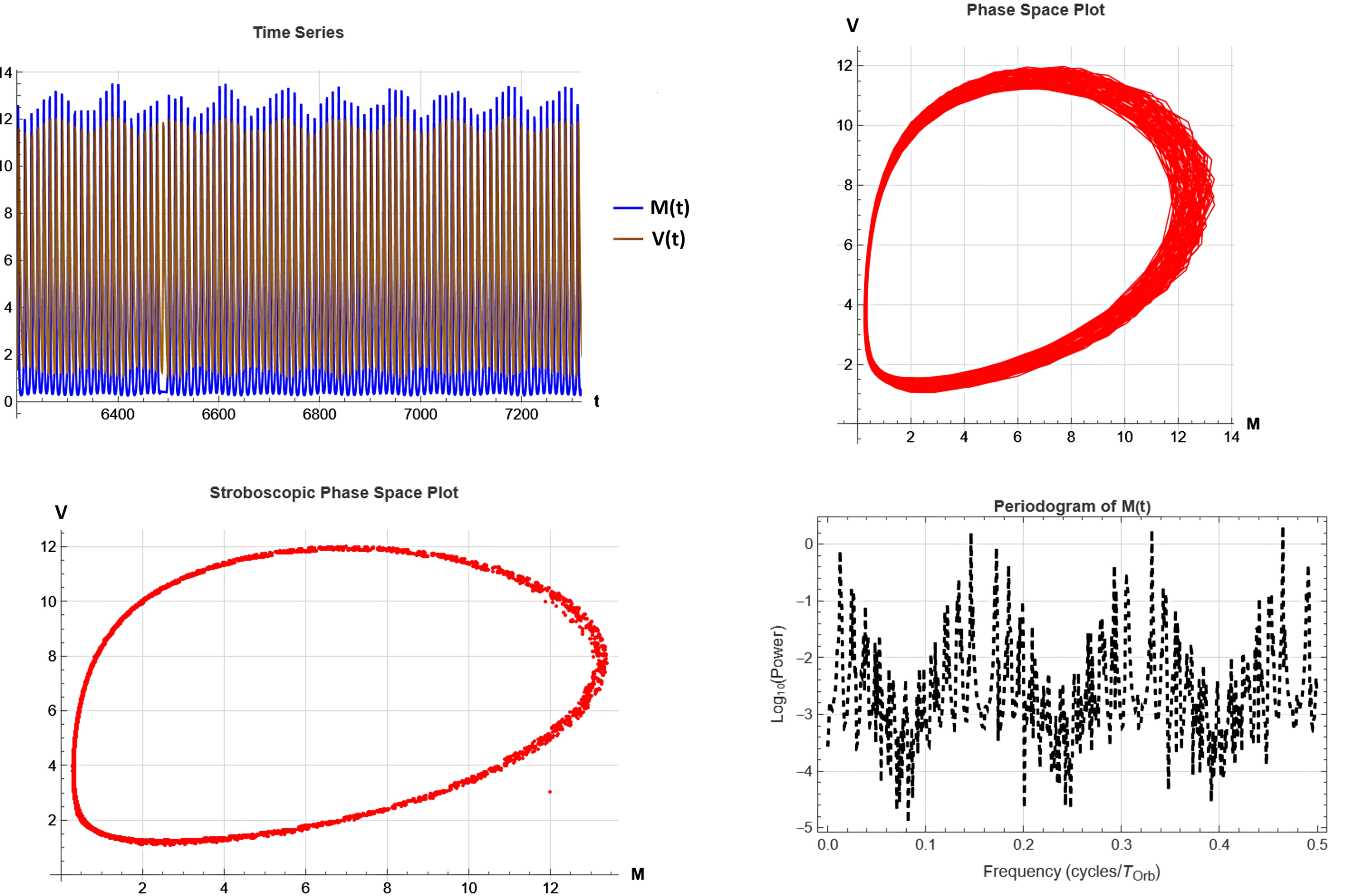}
    \caption{Upper row: time series and phase portrait over a $10\,T_{s,1}$ window. Bottom row: stroboscopic projection sampled every $T_{s,1}$ over a very long window $(10^4 \ T_{s,1})$ and a  periodogram of $M(t)$ over $10\,T_{s,1}$. Parameters from Table~\ref{tab:params} with $(n=1.30, k=0.62)$.}
    \label{fig:Times_Series1}
\end{figure}

\subsection{Time-domain, stroboscopic, and spectral analysis}
\begin{figure}[t]
    \centering
    \includegraphics[width=\linewidth]{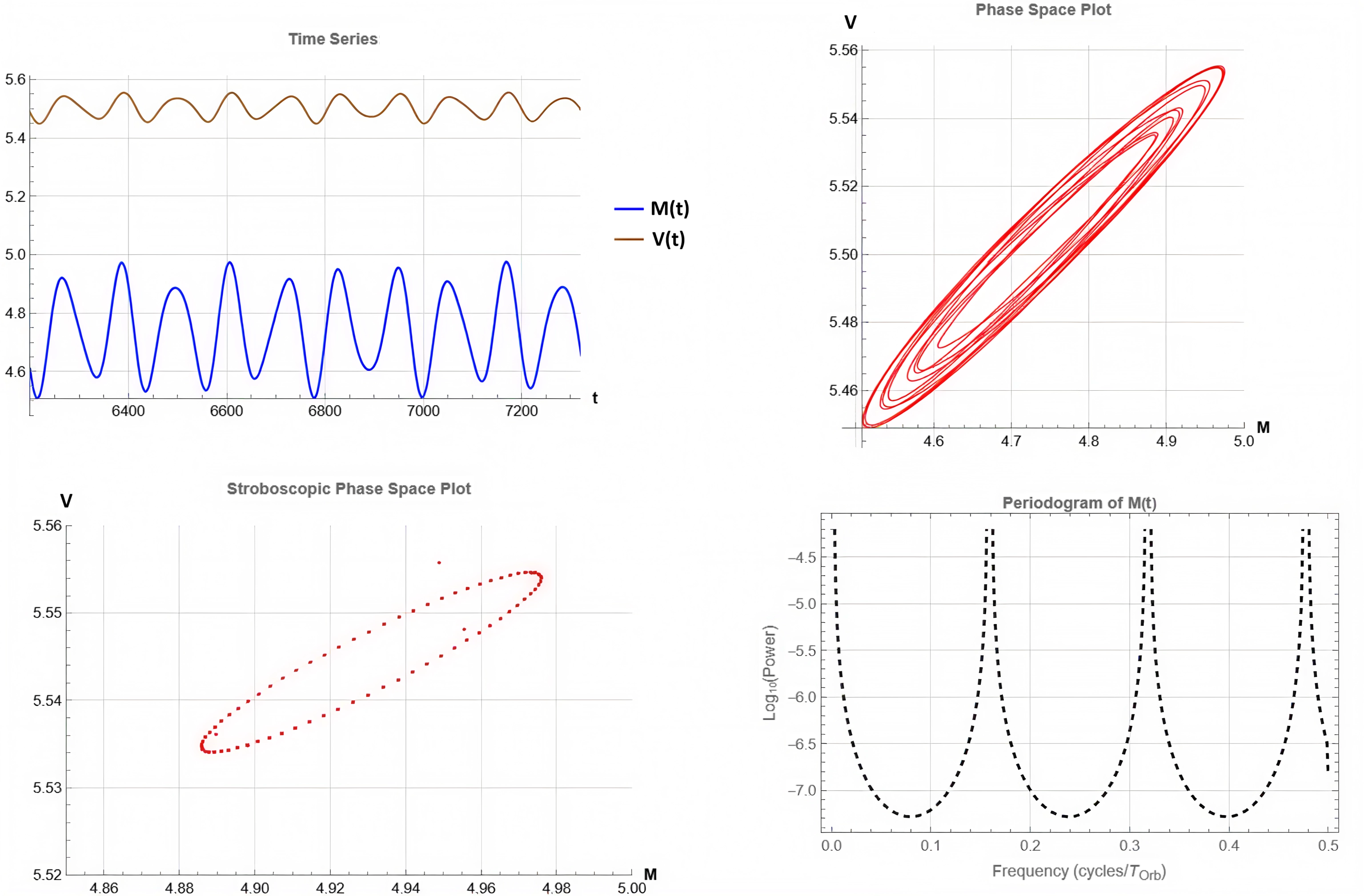}
    \caption{Same set of diagnostics as Fig.~\ref{fig:Times_Series1}. Parameters: $(n=1.28, k=0.54)$.}
    \label{fig:Times_Series2}
\end{figure}

Figures~\ref{fig:Times_Series1} and \ref{fig:Times_Series2} characterize the dynamics using four diagnostics: (i) short time series windows of $M(t)$ and $V(t)$; (ii) the continuous phase portrait $(M(t),V(t))$; (iii) a Prometheus-stroboscopic diagnostic projection sampled once every $T_{s,1}$; and (iv) a one–sided periodogram of $M(t)$. Both figures use identical plotting conventions to enable direct comparison. 

Because the model is driven by two forcing frequencies, sampling once every
Prometheus synodic period does not define a classical Poincaré section of a
singly periodic system. We therefore use the sequence
\[
\{(M(t_k),V(t_k)):t_k=t_0+kT_{s,1}\}
\]
as a Prometheus-stroboscopic diagnostic projection. Its geometry is diagnostically useful: a finite point set indicates frequency locking, a thin closed curve is consistent with quasiperiodic-like bounded motion in the projected variables, while a thickened band or filamented structure indicates strong modulation, intermittent excursions, or proximity to locking.

For spectral analysis, we compute periodograms using uniformly sampled $10\,T_{s,1}$ windows. After subtracting the sample mean and applying a Blackman–Harris taper to reduce leakage, we scale the discrete Fourier transform by the sampling interval and window power ensuring spectral peaks are comparable across parameter regimes. Narrow isolated peaks indicate coherent oscillations; harmonics and combination tones at $m/T_{s,1 }\pm \ell/T_{s,2}$ report nonlinear mixing between the two drives; a broadband floor indicates loss of coherence.

\paragraph*{Figure~\ref{fig:Times_Series1}: $(n=1.30,\;k=0.62)$.}
The $10\,T_{s,1}$ time series show highly non–sinusoidal waveforms with sharp pulses and  cycle asymmetry, reflecting pronounced nonlinearity. The phase portrait traces a thick, non–elliptic loop with cusp-like features and large radial excursions. The stroboscopic projection forms a broad band, indicating persistent modulation  and intermittent angular advance with respect to the \(T_{s,1}\) drive. The periodogram exhibits numerous spectral lines superimposed on an elevated broadband floor  with sidebands around multiples of $1/T_{s,1}$, characteristic of strong nonlinear mixing between the $T_{s,1}$ and $T_{s,2}$ forcings. Overall, these signatures are most consistent with strongly modulated bounded oscillations with a thickened torus-like projection.

\paragraph*{Figure~\ref{fig:Times_Series2}: $(n=1.28,\;k=0.54)$.}
Here the $10\,T_{s,1}$ time series are smooth with only modest amplitude modulation. The phase portrait is nearly elliptical, consistent with motion  in a narrow annulus. The stroboscopic projection forms a thin closed invariant curve, consistent with quasi–periodicity in the projected variables. Consistently, the periodogram shows a small number of sharp, well–separated peaks with minimal broadband background.

\subsection{Unit–circle stroboscopic maps}
To complement the preceding analysis, we use \emph{unit–circle stroboscopic maps} to distinguish angular organization from amplitude variability. For each $(n,k)$ we integrate to $t_{\max}=10^6 \ T_{s,1}$, discard a transient warm–up period, and record strobes at $t_j=t_0+jT_{s,1}$ to form
\[
\mathcal{S}=\{(M(t_j),V(t_j))\}_{j=0}^{N-1}.
\]
Under dual-frequency forcing, this construction should be interpreted as a
stroboscopic projection with respect to the dominant Prometheus period, rather
than as a classical Poincaré section of a singly periodic system. Let $c=\mathrm{mean}(\mathcal{S})$ and we define $z_j=(M(t_j)-c_M)+i(V(t_j)-c_V)$ with radius $r_j=|z_j|$. We then project on the unit circle: $Z_j=z_j/|z_j|=e^{i\theta_j}$. A nearly uniform angular distribution is consistent with smooth quasiperiodic-like motion in this projection; azimuthal clustering or partial arcs suggest proximity to rational rotation numbers or intermittent angular advance, while pre-projection ring thickening is consistent with stronger amplitude modulation. Because the construction depends on the chosen centering and projection, it is used as a diagnostic rather than as a coordinate-invariant classification.

\begin{figure}[]
  \centering
  \includegraphics[width=\linewidth]{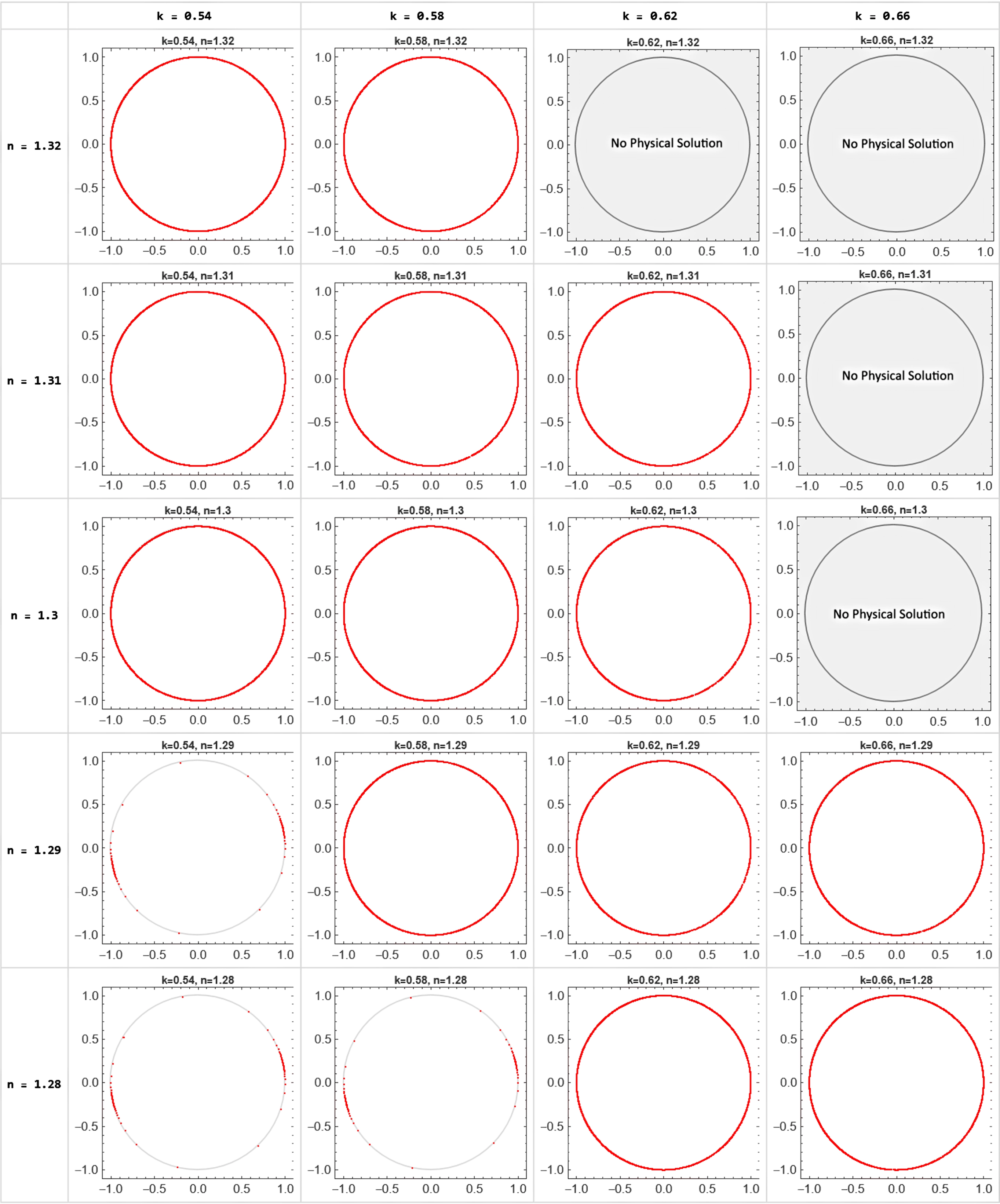}
  \caption{\textbf{Unit–circle stroboscopic maps across $(n,k)$.}
  Each panel shows $Z_j=z_j/|z_j|$ obtained by centering the stroboscopic samples $(M(t_j),V(t_j))$, mapping to the complex plane, and normalizing to the unit circle. }
  \label{fig:fig3}
\end{figure}

Examining Fig.~\ref{fig:fig3} by rows (fixed $n$, increasing $k$) shows two trends. For lower $n$, point distributions nearly uniformly cover the circle but with a non-uniform azimuthal density, consistent with the thin invariant curve  and discrete, low-noise spectral lines in Fig.~\ref{fig:Times_Series2}. As $n$ increases or $k$ strengthens, azimuthal inhomogeneity increases, filling the whole circle and the (pre-projection) ring thickens, reflecting the broadened stroboscopic band and sideband-rich spectra of Fig.~\ref{fig:Times_Series1}. Thus, the unit-circle diagnostic provides a compact geometry-based diagnostic support for the regime classification. For larger combinations of \((n,k)\), the reduced model often fails to sustain physically admissible long-term bounded trajectories and they are labelled as ``No Physical Solution''. They include single-burst collapse, crossing of the physical boundary \(V<0\), and explosive aggregation exceeding the imposed boundedness threshold.
\begin{figure}[]
  \centering
  \includegraphics[width=\linewidth]{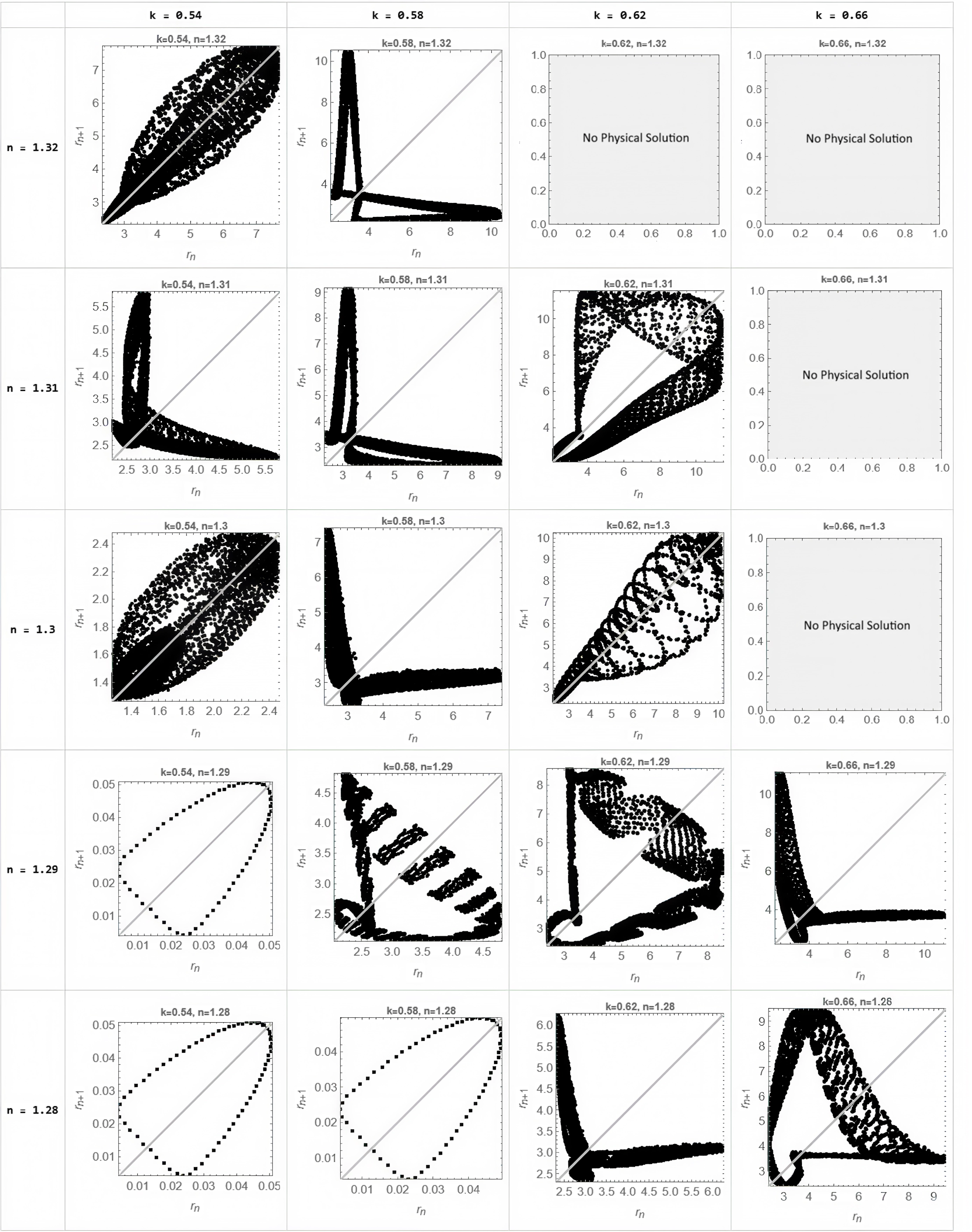}
  \caption{\textbf{Stroboscopic radial return maps across $(n,k)$.} 
  Each panel shows the relation between successive stroboscopic radii $r_{n}$ and $r_{n+1}$ computed from centered samples $(M(t_j),V(t_j))$ taken every $T_{s,1}$. The gray diagonal is $r_{n+1}=r_n$. }
  \label{fig:returnmaps}
\end{figure}

Comparison of the two parameter regimes reveals  the transition from strongly nonlinear, multi–frequency dynamics to coherent quasi–periodicity. At $(n=1.30,k=0.62)$ the dynamics show waveform distortion accompanied by a thick Prometheus-stroboscopic band with features characteristic of modulation and intermittent angular advance. Decreasing to $(1.28,0.54)$ regularizes the dynamics: the Prometheus-stroboscopic set becomes a thin closed circle, exhibiting non-uniform azimuthal point density, consistent with quasiperiodic-like motion in this projection.

\subsection{Stroboscopic radial return maps ($r_{n+1}$ vs.\ $r_n$)}
\label{sec:return-maps}

To isolate amplitude dynamics from angular organization, we construct a radial return map using the same stroboscopic samples as in Fig.~\ref{fig:fig3}. For a given parameter pair $(n,k)$ we integrate Eq.~(1) to a long horizon $t_{\max}= 10^6 \ T_{s,1}$ after discarding transient warm-up segments, then sample the trajectory once per Prometheus period $T_{s,1}$ at times $t_j=t_0+jT_{s,1}$, yielding the set $\mathcal{S}=\{(M(t_j),V(t_j))\}_{j=0}^{N-1}$. The point cloud is centered at its centroid  $c=(\bar M,\bar V)$ and we define the complexified, centered points $z_j=(M(t_j)-\bar M)+i\,(V(t_j)-\bar V)$. 
The stroboscopic radius is
\[
r_j=\lvert z_j\rvert=\sqrt{(M(t_j)-\bar M)^2+(V(t_j)-\bar V)^2}\,,
\]
which measures distance from the centroid in the $(M,V)$ plane and thus serves as a scalar measure of the oscillation amplitude at each stroboscopic sample.  The radial return map plots the successive-pair relation
\[
r_{n} \;\mapsto\; r_{n+1}\,,
\]
together with the identity line $r_{n+1}=r_n$ for reference. Invariant circles with strictly constant radius produce a single point on the diagonal; smooth two–frequency tori with weak amplitude modulation collapse to narrow near-diagonal bands; multi-scale modulation,  intermittent excursions or near locking behavior inflate the cloud off the diagonal and/or produce multi-valued structure. Because the map is one-dimensional in the observable $r$, it directly reports how amplitudes transfer from one strobe to the next, complementing the unit–circle map of Fig.~\ref{fig:fig3} that focuses on angle.

The radial return maps in Fig.~\ref{fig:returnmaps} mirror the progression described on the unit circles of  Fig.~\ref{fig:fig3}. Tight closed clusters near the diagonal indicate nearly constant stroboscopic radius (coherent quasiperiodic-like motion in the projection), whereas broadened, structured clouds signal strong amplitude modulation and intermittent excursions consistent with the thick Prometheus-stroboscopic bands  seen in Figs.~\ref{fig:Times_Series1}–\ref{fig:fig3}.

For the quasiperiodic-like bounded regime of Fig.~\ref{fig:Times_Series2} \big($(n,k)=(1.28,0.54)$\big),  the corresponding $r_{n+1}$ vs.\ $r_n$ panel collapses to a tight swarm straddling the diagonal, reflecting an almost constant stroboscopic radius with only weak, smooth amplitude transfer between successive strobes. In other words, the amplitude is slaved to a gentle two–frequency modulation and exhibits no visible evidence, within this diagnostic, of strongly multi-valued return behavior.

At the nonlinear setting of Fig.~\ref{fig:Times_Series1} \big($(n,k)=(1.3,0.62)$\big), the  radial return panel expands into a visibly thicker cloud about the diagonal, signalling stronger amplitude memory from strobe to strobe and intermittent departures from a quasi–one–dimensional closed curve. In this regime, the return relation is no longer nearly single-valued: successive amplitudes explore a wider interval, in line with the modulation and intermittent excursions  inferred from the other diagnostics.

% ---------- Rotation-number heatmap (single panel) ----------
\begin{figure}[t]
  \centering
  % replace filename as needed
  \includegraphics[width=1.1\linewidth]{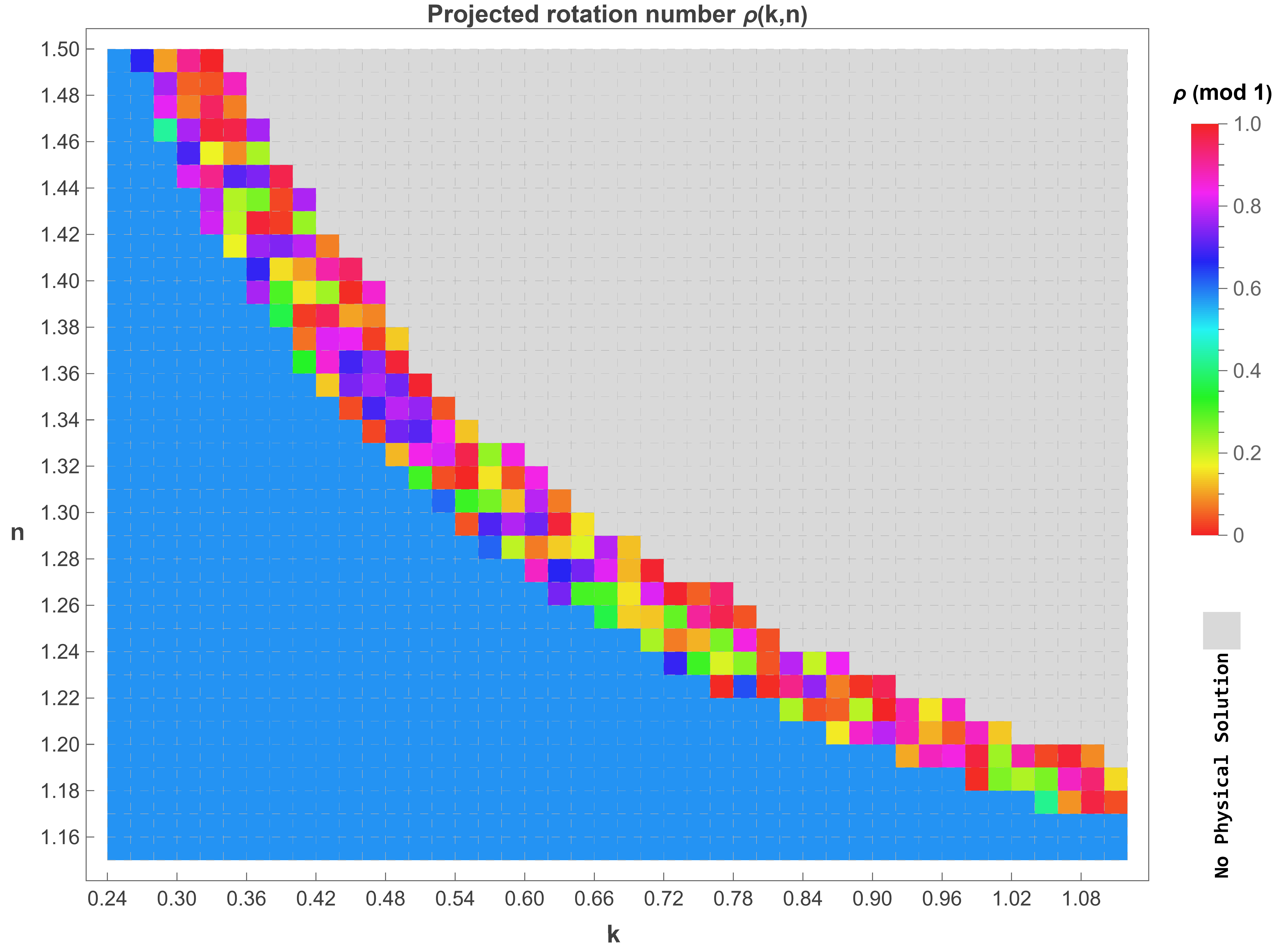}
  \caption{\textbf{Rotation number heatmap over $(k,n)$.}
Color encodes \(\rho(k,n)\bmod 1\), computed from centered Prometheus-stroboscopic samples, using a cyclic color map so that \(\rho=0\) and \(\rho=1\) are represented by
the same color. Grey cells denote parameter pairs for which the reduced model does not produce a physically admissible long-term bounded trajectory.}
  \label{fig:rhoHeat}
\end{figure}

\subsection{Rotation-number heatmap}

For each point on a uniformly spaced parameter grid \((k,n)\), with
\(k\in[0.24,1.10]\) and \(n\in[1.16,1.50]\), we integrate the dual-forcing
system. After discarding initial transients, we sample a stroboscopic set once
per Prometheus synodic period \(T_{s,1}\). Because the system is driven by two
forcing frequencies, this sampling is not a classical Poincar\'e section of a
singly periodic system. It is instead a diagnostic stroboscopic projection with
respect to the dominant Prometheus forcing:
\[
\mathcal{S}=\{(M(t_j),V(t_j)):\ t_j=t_0+jT_{s,1},\ j=0,\dots,N-1\}.
\]
We center \(\mathcal{S}\) at its centroid \(c=(c_M,c_V)\) and form the complex,
centered sequence
\[
z_j=(M(t_j)-c_M)+i(V(t_j)-c_V),\qquad r_j=|z_j|.
\]
The angular increments
\[
\Delta\theta_j=\operatorname{wrap}_{(-\pi,\pi]}
\big(\arg z_{j+1}-\arg z_j\big)
\]
define the projected stroboscopic rotation number
\[
\rho(k,n)=\frac{1}{2\pi}\,\frac{1}{N-1}\sum_j \Delta\theta_j .
\]
In Fig.~\ref{fig:rhoHeat}, we plot the fractional part
\[
\rho_{\mathrm{frac}}(k,n)
=
\rho(k,n)-\lfloor \rho(k,n)\rfloor
=
\rho(k,n)\bmod 1,
\]
so that all values lie in \([0,1)\). A cyclic colour map is used, so that
\(\rho_{\mathrm{frac}}=0\) and \(\rho_{\mathrm{frac}}=1\) are represented by the same colour and values near these endpoints appear adjacent.  Because the phase is extracted from centered coordinates in the \((M,V)\) plane, this rotation number is projection-dependent. It is therefore used as a compact diagnostic for comparing parameter regimes, not as a coordinate-invariant dynamical invariant of the full dual-frequency forced system. 

Parameter combinations failing admissibility or reliability checks are marked as
``No Physical Solution'' and shaded grey. These include trajectories that cross
\(M<0\) or \(V<0\), collapse to a near-zero mass state, yield too few valid
post-transient stroboscopic samples, or exceed the imposed boundedness
threshold. As discussed above, this label denotes loss of physically admissible
long-term bounded dynamics in the reduced model, not literal nonexistence of a mathematical ODE solution.

Smooth spatial variation of the projected rotation number,
when combined with thin stroboscopic loops and discrete spectra, is interpreted as evidence consistent with quasiperiodic-like bounded motion. Abrupt colour changes, narrow bands, or plateau-like regions are interpreted as indicators of  proximity to locking or stronger modulation.

The rotation-number map in Fig.~\ref{fig:rhoHeat} is consistent with the
behaviour seen in Figs.~\ref{fig:Times_Series1}--\ref{fig:Times_Series2}, the
unit-circle stroboscopic panels in Fig.~\ref{fig:fig3}, and the radial return
maps in Fig.~\ref{fig:returnmaps}. At \((n,k)=(1.28,0.54)\), the time series are
smooth and the stroboscopic projection collapses to a thin loop; in
Fig.~\ref{fig:rhoHeat}, this point lies in a smoothly varying bounded region of
the projected rotation-number map. At \((n,k)=(1.30,0.62)\), the pulses and
broadened stroboscopic band indicate stronger modulation and proximity to
near-locking; consistently, Fig.~\ref{fig:rhoHeat} places this point within a
region of sharper colour variation near the transition to non-admissible
dynamics.

In analogy with the radial return map, tight clusters near the diagonal
\(r_{j+1}\approx r_j\) indicate weak amplitude modulation and a nearly constant
radius of the stroboscopic loop. These cases correspond to quasiperiodic-like
bounded zones in Fig.~\ref{fig:rhoHeat}. As \(k\) or \(n\) increase and the
heatmap develops sharper colour changes or narrow transition bands, the return
clouds broaden, tilt, or form multi-band structures, indicating stronger
cycle-to-cycle amplitude variation.

\subsection{Discussion}
Our analysis shows that the reduced dual-forcing model can produce
self-organized bounded aggregation-fragmentation oscillations on timescales
relevant to moon-ring forcing. By integrating time-domain, spectral, and geometric diagnostics, we establish a consistent physical interpretation: Prometheus acts as the dominant driver sustaining clumping--dispersal cycles, while Pandora introduces a weaker secondary frequency that can generate slower amplitude modulation and quasiperiodic-like envelopes in the reduced model. This dynamical duality mirrors the interplay between resonant confinement and weak perturbative diffusion inferred from Cassini observations, indicating that deterministic dual-frequency modulation may contribute to F-ring variability alongside stochastic impacts, spatially extended resonant dynamics, and collisional processes. 

Central to the model is the nonlinear aggregation law $M^n$, which governs how efficiently existing clumps capture additional material. When $n$ exceeds unity, aggregation accelerates faster than linear coagulation, creating a positive feedback where larger aggregates have enhanced collision cross-sections. This leads to superlinear growth that can trigger large-amplitude aggregation bursts. Our numerical diagnostics reveal that increasing $n$ or the aggregation efficiency $k$ drives the system from regular quasiperiodic-like projected motion toward strongly modulated and intermittent states, manifested as broadened Prometheus-stroboscopic bands and sideband-rich spectra. Physically, this corresponds to a regime where collisions and self-gravity reinforce moon-driven compression, producing transient over-densities that exceed the fragmentation threshold before relaxing. The alternation between buildup and dispersal constitutes a self-sustained oscillation whose amplitude is governed by the competition between nonlinear growth and dissipative fragmentation.

The appearance of thickened toroidal structures and rotation-number plateaus is consistent with stronger amplitude-phase coupling between the Prometheus and Pandora drives. In planetary terms, this reflects how nearby moons, through their slightly incommensurate synodic frequencies, generate beating patterns that periodically strengthen and weaken the ring's azimuthal structure. The rotation-number map displays smooth quasiperiodic-like regions separated by near-locking bands, analogous to Arnold tongues in forced oscillators. Within these tongues, the projected dynamics approach near-rational frequency relations, providing a possible reduced-model mechanism for episodic strengthening and weakening of aggregate proxies. Between these tongues, phase drift resumes and the system explores an entire torus-like set, corresponding to the observed alternation between well-defined strands and diffuse textures.

The transition between the representative regimes $(n, k) = (1.28, 0.54)$ and $(n, k) = (1.30, 0.62)$ demonstrates remarkable structural instability, where a $\sim 1.5\%$ variation in $n$ produces qualitatively different dynamics. The lower parameter pair yields smooth quasi-periodicity, indicating that aggregation and fragmentation remain in balance, while the higher pair leads to pronounced modulation and intermittent excursions characteristic of nonlinear resonance overlap. This sensitivity suggests that the reduced model operates near a boundary between smooth bounded modulation and stronger large-amplitude or near-locking behavior. A precise bifurcation classification would require continuation or Lyapunov-spectrum analysis and is left for future work. The existence of bounded but non-periodic oscillations suggests a mechanism by which bounded aggregate variability could be maintained.

Notably, the model achieves considerable descriptive power with minimal complexity: with only two dynamical variables and a compact set of physically interpretable parameters, it reproduces several qualitative dynamical signatures relevant to reduced
descriptions of F-ring clump recurrence, envelope modulation, and quasi-stability over thousands of orbital periods. The spectral diagnostics confirm that dual forcing introduces combination tones consistent with nonlinear mixing, while the geometric maps are consistent with torus-like projected trajectories in the bounded regimes. Consequently, part of the apparent variability may be interpretable through deterministic multi-frequency modulation.

The dynamical regimes we identify here, from quasiperiodic-like regimes to modulated oscillations, represent universal responses of nonlinear oscillators to multi-frequency forcing. Our system exhibits behavior analogous to other driven dissipative systems, such as forced Josephson junctions \cite{meister_resonators_2015, baxevanis_inductively_2025}, where the interplay between nonlinearity and multiple driving frequencies produces rich synchronization structures. This suggests that reduced models of the F ring can provide a useful setting for studying general principles of multi-frequency dynamics in astrophysical contexts, with implications for other granular systems subject to multiple periodic perturbations.

\subsection{Limitations}
Although the dual–moon model captures qualitative dynamical signatures relevant to F-ring aggregation-fragmentation, it remains a reduced mean-field representation that omits several layers of physical complexity. The formulation assumes spatial homogeneity and neglects azimuthal phase gradients and stochastic impacts from meteoroid fluxes, which may locally alter collision rates and velocity dispersion. The aggregation–fragmentation law is parameterized phenomenologically, and the exponent $n$ absorbs multiple microphysical processes that could vary with particle size and composition. 

The present study does not perform a direct fit to Cassini imaging, RSS, UVIS,
or VIMS occultation profiles. The comparison with observations is therefore made
at the level of physically motivated timescales and qualitative dynamical
signatures, not at the level of measured clump brightness, optical depth, or
aggregate-size distributions. 
To make this observational status explicit, Appendix~\ref{app:observational-predictions},
Table~\ref{tab:observational_predictions} summarizes the main model-derived predictions,
possible Cassini data counterparts, and whether each item is already observationally
constrained or remains a prediction for future testing.

Furthermore, the two–frequency forcing is prescribed deterministically, without feedback from the evolving ring mass distribution onto the moons’ orbital elements and without accounting for time-delayed effects. The stroboscopic maps, unit-circle projections, radial return maps, and projected rotation numbers are diagnostic tools. They do not by themselves constitute a full bifurcation or chaos analysis. Including finite response delays between aggregation, stirring, and fragmentation
could enlarge the effective phase space and may generate dynamical regimes not
captured by the present two-variable ODE model. Future extensions should embed this reduced model within radially resolved or $N$-body frameworks to test the persistence of the quasiperiodic-like regimes under more realistic collisional and gravitational coupling.

%=============================================================
\section{Conclusions}

Our study develops and analyzes a minimal nonlinear framework linking
aggregation-fragmentation feedback to oscillatory dynamics under dual-moon
forcing. By extending the predator--prey formulation to include Prometheus and
Pandora as distinct prescribed periodic drivers, the model provides a reduced
dynamical mechanism relevant to several Cassini-motivated features, including
recurrent clumping, modulation envelopes, and long-lived variability in the
F-ring environment. Within this reduced system, nonlinear aggregation combined
with dual-frequency forcing can generate bounded, self-sustained oscillations
whose projected character changes from quasiperiodic-like modulation
\((n=1.28,k=0.54)\) to strongly modulated, near-locking behavior
\((n=1.30,k=0.62)\) as nonlinear growth is strengthened. These dynamics
correspond, at the level of the model variables, to alternating episodes of
mass concentration and dispersal governed by the interaction between
aggregation, fragmentation, and moon-driven forcing. The comparison between the
two representative regimes shows how small parameter changes in the reduced
system can produce qualitatively different diagnostic signatures.

From a broader perspective, the model suggests that a boundary between bounded
modulation and near-locking behavior may be dynamically relevant for reduced
descriptions of F-ring variability. Deterministic modulation should therefore
be viewed as complementary to stochastic impacts, spatially extended resonant
dynamics, and collisional processes, rather than as a replacement for them. In
the bounded regimes, perturbations are accommodated through amplitude and phase
adjustment within the model, whereas stronger nonlinear aggregation can drive
the system outside the physically admissible long-term regime. Beyond its
planetary context, the framework provides a starting point for studying
multi-frequency forcing in driven dissipative granular systems. Future work
should incorporate spatial structure, stochastic perturbations, time-delayed
responses, and direct comparison with Cassini imaging and occultation
diagnostics, including clump brightness, optical-depth variation, and aggregate
detection statistics. In particular, the model predicts that aggregate-mass or brightness proxies should exhibit modulation and sideband structure associated with the
Prometheus-Pandora forcing and beat timescales, a prediction that can be tested
against future analyses of Cassini-derived clump, brightness, and occultation data.

%=============================================================
\clearpage
\begin{appendices}
\section{Derivation and scaling of the dual-moon aggregation-fragmentation model}
\label{app:model-derivation}

This appendix gives the modelling assumptions leading to Eq.~(1). The formulation builds on the predator--prey description of moon-triggered clumping in Saturn's rings developed by Esposito et al.~\cite{esposito_predatorprey_2012}, in which the two state variables are the mean aggregate mass and the squared relative velocity dispersion. We retain this two-variable mean-field structure, but extend it by including a nonlinear aggregation term and an explicit two-frequency forcing associated with Prometheus and Pandora.

Let $  M(t)$ denote the mean aggregate mass of the local particle ensemble, and let $V(t) = v_{\rm rel}^2(t)$ denote the squared relative velocity dispersion of the particles.  All times in the numerical model are measured in units of the local F-ring orbital period \(T_{\rm orb}\). Thus, in dimensionless units,
\begin{equation}
    T_{\rm orb}=1.
\end{equation}
When a dimensional interpretation is required, the F-ring orbital period at radius \(a\) is $T_{\rm orb}^{\rm phys}
    =
    2\pi \sqrt{\frac{a^3}{\mu_{\rm S}}}$ , where \(\mu_{\rm S}\) is Saturn's gravitational parameter. For the F-ring core, this gives approximately  $T_{\rm orb}^{\rm phys} \simeq 0.620\ {\rm days}$.   The value \(T_{\rm orb}=1\) used in the simulations is therefore a nondimensional unit of time, while \(0.620\) days is used only to convert model times back into physical units.

The mean collision time is approximated by $T = T_{\rm coll} = \frac{T_{\rm orb}}{4\tau}$, where \(\tau\) is the local optical depth. With \(\tau=0.1\), this gives  $T  = 2.5$ in units of \(T_{\rm orb}\). 

\subsection{Mass growth and fragmentation}

The starting point is the mass equation in the predator--prey model,
\begin{equation}
    \frac{dM}{dt}
    =
    \frac{M}{T_{\rm acc}}
    -
    \frac{v_{\rm rel}^2}{v_{\rm th}^2}
    \frac{M}{T_{\rm coll}},
    \label{eq:esposito-mass}
\end{equation}
where \(T_{\rm acc}\) is the characteristic accretion or aggregation time, \(T_{\rm coll}\) is the collision time, and \(v_{\rm th}\) is the threshold velocity above which collisions tend to disrupt aggregates rather than promote sticking.  In the reduced model (proposed in \cite{esposito_predatorprey_2012}),  we have $ T_{\rm acc}=T_{\rm coll}=T$, which gives
\begin{equation}
    \frac{dM}{dt}
    =
    \frac{M}{T}
    -
    \frac{V}{v_{\rm th}^2}
    \frac{M}{T}.
    \label{eq:linear-mass}
\end{equation}
The first term represents local aggregate growth. The second term represents fragmentation or erosion when the squared relative velocity dispersion \(V\) becomes large compared with \(v_{\rm th}^2\). To account phenomenologically for enhanced mass-dependent aggregation caused by effects such as self-gravity, shear reversal, gravitational compression, or the increased cross-section of larger clumps, we add a nonlinear aggregation term $ kM^n$. The parameter \(n\) controls the nonlinearity of aggregate growth, while \(k\) controls the strength of this additional aggregation pathway. This gives
\begin{equation}
    \frac{dM}{dt}
    =
    \frac{M}{T}
    -
    \frac{V}{v_{\rm th}^2}
    \frac{M}{T}
    +
    kM^n.
    \label{eq:mass-final}
\end{equation}
In this paper \(k\) and \(n>1\) are treated as phenomenological parameters. The parameter \(k\) is not inferred directly from microphysical sticking laws; instead, it is used to explore weak to strong nonlinear aggregation regimes within the reduced mean-field model.

\subsection{Velocity-dispersion equation}

The second equation describes the evolution of
\begin{equation}
    V(t)=v_{\rm rel}^2(t),
\end{equation}
where $v_{\rm rel}$ is the relative velocity dispersion. Following ~\cite{esposito_predatorprey_2012}, the velocity-dispersion equation contains two baseline terms: collisional dissipation and gravitational/viscous stirring by aggregates.

The dissipative contribution is
\begin{equation}
    -\frac{1-\epsilon^2}{T}V,
    \label{eq:dissipation}
\end{equation}
where \(\epsilon\) is the coefficient of restitution. This term is linear in \(V=v_{\rm rel}^2\) because inelastic collisions remove a fixed fraction \(1-\epsilon^2\) of the relative kinetic energy per collision. Since relative kinetic energy is proportional to \(v_{\rm rel}^2\), the damping law is naturally linear in \(V\), not in \(v_{\rm rel}\).

The stirring term is written as
\begin{equation}
    \left(\frac{M}{M_0}\right)^2
    \frac{v_{\rm esc}^2}{T_{\rm orb}},
    \label{eq:stirring}
\end{equation}
where \(M_0\) is a reference aggregate mass and \(v_{\rm esc}\) is the escape velocity from the surface of an aggregate of mass \(M_0\). This term approximates the increase in velocity dispersion caused by the gravitational stirring of the particle ensemble by local aggregates. The factor \((M/M_0)^2\) expresses the stronger gravitational stirring caused by larger aggregates relative to the reference mass.

For a reference aggregate of mass \(M_0\) and characteristic radius \(R_0\), the escape velocity is
\begin{equation}
    v_{\rm esc}
    =
    \sqrt{\frac{2GM_0}{R_0}}.
    \label{eq:vesc}
\end{equation}
If the aggregate is approximated as a sphere of internal density \(\rho\), then $M_0 = \frac{4\pi}{3}\rho R_0^3$. In the parameter values used here, \(M_0=2\times 10^9\,{\rm g}\) corresponds to a roughly \(10\) m icy aggregate with density \(\rho\simeq0.5\,{\rm g\,cm^{-3}}\), and \(v_{\rm esc}\simeq0.5\,{\rm m\,s^{-1}}\). 

Combining dissipation and stirring gives the unforced velocity equation
\begin{equation}
    \frac{dV}{dt}
    =
    -
    \frac{1-\epsilon^2}{T}V
    +
    \left(\frac{M}{M_0}\right)^2
    \frac{v_{\rm esc}^2}{T_{\rm orb}}.
    \label{eq:velocity-unforced}
\end{equation}

\subsection{Moon forcing and sign convention}

The moon perturbation modifies the velocity-dispersion equation because the perturbation changes local streamline crowding, relative velocity, and stirring. For a single moon with forcing frequency \(\omega\), the forced velocity equation may be written in the form
\begin{equation}
    \frac{dV}{dt}
    =
    -
    \frac{1-\epsilon^2}{T}V
    +
    \left(\frac{M}{M_0}\right)^2
    \frac{v_{\rm esc}^2}{T_{\rm orb}}
    -
    A_0\cos(\omega t).
    \label{eq:single-forced}
\end{equation}
The negative sign is a phase convention. In the present interpretation, the phase \(\omega t=0\) corresponds to the part of the moon encounter that promotes streamline crowding and reduces the relative velocity dispersion. A reduction in \(V\) promotes aggregation through Eq.~\eqref{eq:mass-final}. Half a forcing cycle later, the sign of the cosine reverses and the forcing contributes to increased velocity dispersion. Thus the term should not be interpreted as the moon always reducing \(V\); rather, it represents the oscillatory compression--stirring effect of the moon encounter with a chosen phase origin.

Equivalently, a different choice of phase origin would write the same forcing as $A_0\cos(\omega t+\phi)$, with a different phase \(\phi\). The sign in Eq.~\eqref{eq:single-forced} is therefore not a separate physical assumption, but a convention that fixes the reference phase of the forcing relative to the aggregation cycle.

The F ring is affected primarily by Prometheus, with Pandora providing a weaker secondary perturbation. We therefore replace the single forcing term by a two-frequency term,
\begin{equation}
    F(t)
    =
    A_0\cos(\omega_1 t)
    +
    A_0\eta\cos(\omega_2 t),
    \label{eq:dual-forcing}
\end{equation}
where \(A_0\) is the Prometheus forcing amplitude, \(\eta\in[0,1]\) is the relative strength of the Pandora forcing, and $\omega_i = \frac{2\pi}{T_{s,i}}$.  Here \(T_{s,1}\) is the Prometheus-F-ring synodic period and \(T_{s,2}\) is the Pandora--F-ring synodic period.

\subsection{Full model}

Combining Eqs.~\eqref{eq:mass-final}, \eqref{eq:velocity-unforced}, and \eqref{eq:dual-forcing}, the model used in the main text is
\begin{align}
    \frac{dM}{dt}
    &=
    \frac{M}{T}
    -
    \frac{V}{v_{\rm th}^2}
    \frac{M}{T}
    +
    kM^n,
    \label{eq:appendix-final-M}
    \\
    \frac{dV}{dt}
    &=
    -
    \frac{1-\epsilon^2}{T}V
    +
    \left(\frac{M}{M_0}\right)^2
    \frac{v_{\rm esc}^2}{T_{\rm orb}}
    -
    \left[
    A_0\cos(\omega_1 t)
    +
    A_0\eta\cos(\omega_2 t)
    \right].
    \label{eq:appendix-final-V}
\end{align}
This is Eq.~(1) in the main text.

\subsection{Physically admissible regimes}

Because \(M(t)\) is a mass proxy and \(V(t)=v_{\rm rel}^2(t)\) is a squared velocity dispersion, physically admissible solutions must satisfy
\begin{equation}
    M(t)\geq 0,
    \qquad
    V(t)\geq 0.
\end{equation}
Parameter combinations for which the numerical trajectory crosses \(M<0\), crosses \(V<0\), collapses to a near-zero mass state, or exceeds an imposed boundedness threshold (possibly diverging solutions) are not interpreted as Saturn-like long-term ring dynamics. In the parameter maps these cases are therefore labelled as ``No Physical Solution'', rather than as a literal nonexistence of a mathematical solution. This distinction is important because the reduced two-variable model may leave its physical range even though the formal ordinary differential equation can still be continued mathematically.

\section{Numerical robustness checks}
\label{app:numerical-robustness}

To test the robustness of the numerical regime classification, representative
parameter values were recomputed under three increasingly strict numerical
settings. The baseline setting was
\texttt{WorkingPrecision -> MachinePrecision},
\texttt{AccuracyGoal -> 10}, \texttt{PrecisionGoal -> 10}, and
\texttt{MaxStepFraction -> 1/50}. The intermediate setting used
\texttt{WorkingPrecision -> 30}, \texttt{AccuracyGoal -> 15},
\texttt{PrecisionGoal -> 15}, and \texttt{MaxStepFraction -> 1/75}. The refined
setting used \texttt{WorkingPrecision -> 40}, \texttt{AccuracyGoal -> 20},
\texttt{PrecisionGoal -> 20}, and \texttt{MaxStepFraction -> 1/100}. In all
cases, \texttt{MaxSteps -> Infinity} was used.

The two representative bounded regimes used in the main figures were robust
under these refinements. The quasiperiodic-like case \((n,k)=(1.28,0.54)\)

retained its smooth time series, thin Prometheus-stroboscopic curve, and
discrete spectral peaks. The strongly modulated case \((n,k)=(1.30,0.62)\)
retained its broadened stroboscopic band, large-amplitude modulation, and
sideband-rich spectral structure. Parameter values classified as ``No Physical
Solution'' also remained outside the physically admissible long-term regime:
they either failed to produce a valid long-term output or reached oscillations
crossing the physical boundary \(V=0\).

\begin{table}[h]
\centering
\caption{Numerical robustness checks under increasing precision and smaller
maximum step fraction.}
\label{tab:numerical-robustness}
\begin{tabular}{llll}
\hline
\((n,k)\) & Baseline setting & Intermediate setting & Refined setting \\
\hline
\((1.28,0.54)\) & quasiperiodic-like bounded & quasiperiodic-like bounded & quasiperiodic-like bounded \\
\((1.30,0.62)\) & strongly modulated bounded & strongly modulated bounded & strongly modulated bounded \\
\((1.40,0.75)\) & oscillations with \(V<0\) & oscillations with \(V<0\) & oscillations with \(V<0\) \\
\((1.50,0.65)\) & no valid output & oscillations with \(V<0\) & oscillations with \(V<0\) \\
\((2.00,0.50)\) & no valid output & no valid output & no valid output \\
\hline
\end{tabular}
\end{table}

These checks support interpreting the main bounded regimes as numerically
robust and the grey parameter regions as loss of physically admissible
long-term bounded dyn

amics in the reduced model, rather than as artifacts of a
single numerical setting. For non-admissible cases, the exact stopping time can
depend on the imposed admissibility threshold, but the qualitative
classification is unchanged.

\section{Observational Predictions and Data Limitations}
\label{app:observational-predictions}

The present model is not fitted directly to Cassini imaging, RSS, UVIS, or VIMS
occultation profiles. Nevertheless, several outputs of the reduced dual-moon
model correspond to quantities that could be searched for in Cassini-derived
data products. Table~\ref{tab:observational_predictions} summarizes the main
observational links suggested by the model. The table is intended as an
observational roadmap rather than a validation claim: the beat period and
sideband spacing are new model predictions for future testing, while the
aggregate-mass scale provides a broad comparison with existing Cassini-based
constraints.
\begin{table}[htbp]
\centering
\caption{Observational predictions and validations.}
\label{tab:observational_predictions}
\footnotesize
\renewcommand{\arraystretch}{1.05}
\setlength{\tabcolsep}{4pt}
\begin{tabular}{llll}
\hline
Quantity & Predicted value & Observable proxy& Validation \\
\hline

\makecell[l]{Beat period} &
\makecell[l]{
\(T_{\mathrm{beat}}\approx120\) days
} &
\makecell[l]{Brightness envelopes/ ISS images \cite{french2024saturn}} &
\makecell[l]{Prediction to be tested} \\

\makecell[l]{Frequency spacing} &
\makecell[l]{
\(\pm 1/T_{s,2}\) relative to \(1/T_{s,1}\)
} &
\makecell[l]{Secondary peaks/ ISS data} &
\makecell[l]{Prediction to be tested} \\

\makecell[l]{Mass range} &
\makecell[l]{\(M\sim0.5\text{--}12\times10^{9}\,\mathrm{g}\)}&
\makecell[l]{UVIS  occultation profiles} &
\makecell[l]{Consistent with \cite{esposito_predatorprey_2012, cuzzi2024saturn}}\\

\hline
\end{tabular}
\end{table}

\end{appendices}

\bibliography{satbiblio}

@article{baxevanis_inductively_2025,
	title = {Inductively coupled {Josephson} junctions: {A} platform for rich neuromorphic dynamics},
	volume = {111},
	issn = {2470-0045, 2470-0053},
	shorttitle = {Inductively coupled {Josephson} junctions},
	url = {https://link.aps.org/doi/10.1103/PhysRevE.111.044214},
	doi = {10.1103/PhysRevE.111.044214},
	language = {en},
	number = {4},
	journal = {Physical Review E},
	author = {Baxevanis, G. and Hizanidis, J.},
	month = apr,
	year = {2025},
	pages = {044214},
}

@article{meister_resonators_2015,
	title = {Resonators coupled to voltage-biased {Josephson} junctions: {From} linear response to strongly driven nonlinear oscillations},
	volume = {92},
	copyright = {http://link.aps.org/licenses/aps-default-license},
	issn = {1098-0121, 1550-235X},
	shorttitle = {Resonators coupled to voltage-biased {Josephson} junctions},
	url = {https://link.aps.org/doi/10.1103/PhysRevB.92.174532},
	doi = {10.1103/PhysRevB.92.174532},
	language = {en},
	number = {17},
	journal = {Physical Review B},
	author = {Meister, S. and Mecklenburg, M. and Gramich, V. and Stockburger, J. T. and Ankerhold, J. and Kubala, B.},
	month = nov,
	year = {2015},
	pages = {174532},
}

@article{ross_predator-prey_2016,
	title = {Predator-prey dynamics stabilised by nonlinearity explain oscillations in dust-forming plasmas},
	volume = {6},
	copyright = {2016 The Author(s)},
	issn = {2045-2322},
	url = {https://www.nature.com/articles/srep24040},
	doi = {10.1038/srep24040},
	language = {en},
	number = {1},
	journal = {Scientific Reports},
	author = {Ross, A. E. and McKenzie, D. R.},
	month = apr,
	year = {2016},
	pages = {24040},
}

@book{nayfeh2024nonlinear,
  title={Nonlinear oscillations},
  author={Nayfeh, Ali H and Mook, Dean T},
  year={2024},
  publisher={John Wiley \& Sons},
 address   = {New York},
}

@incollection{colwell_structure_2009,
	address = {Dordrecht},
	title = {The {Structure} of {Saturn}'s {Rings}},
	isbn = {9781402092176},
	booktitle = {Saturn from {Cassini}-{Huygens}},
	publisher = {Springer Netherlands},
	author = {Colwell, J. E. and Nicholson, P. D. and Tiscareno, M. S. and Murray, C. D. and French, R. G. and Marouf, E. A.},
	editor = {Dougherty, Michele K. and Esposito, Larry W. and Krimigis, Stamatios M.},
	year = {2009},
	doi = {10.1007/978-1-4020-9217-6\_13},
	pages = {375--412},
}

@article{tiscareno_analytic_2010,
	title = {{AN} {ANALYTIC} {PARAMETERIZATION} {OF} {SELF}-{GRAVITY} {WAKES} {IN} {SATURN}'{S} {RINGS}, {WITH} {APPLICATION} {TO} {OCCULTATIONS} {AND} {PROPELLERS}},
	volume = {139},
	issn = {0004-6256, 1538-3881},
	url = {https://iopscience.iop.org/article/10.1088/0004-6256/139/2/492},
	doi = {10.1088/0004-6256/139/2/492},
	number = {2},
	journal = {The Astronomical Journal},
	author = {Tiscareno, Matthew S. and Perrine, Randall P. and Richardson, Derek C. and Hedman, Matthew M. and Weiss, John W. and Porco, Carolyn C. and Burns, Joseph A.},
	month = feb,
	year = {2010},
	pages = {492--503},
}

@article{colwell_self-gravity_2007,
	title = {Self-gravity wakes and radial structure of {Saturn}'s {B} ring},
	volume = {190},
	issn = {0019-1035},
	url = {https://www.sciencedirect.com/science/article/pii/S0019103507001030},
	doi = {10.1016/j.icarus.2007.03.018},
	number = {1},
	journal = {Icarus},
	author = {Colwell, J. E. and Esposito, L. W. and Sremčević, M. and Stewart, G. R. and McClintock, W. E.},
	month = sep,
	year = {2007},
	pages = {127--144},
}

@article{nicholson_self-gravity_2010,
	series = {Cassini at {Saturn}},
	title = {Self-gravity wake parameters in {Saturn}’s {A} and {B} rings},
	volume = {206},
	issn = {0019-1035},
	url = {https://www.sciencedirect.com/science/article/pii/S0019103509003182},
	doi = {10.1016/j.icarus.2009.07.028},
	number = {2},
	journal = {Icarus},
	author = {Nicholson, P. D. and Hedman, M. M.},
	month = apr,
	year = {2010},
	pages = {410--423},
}

@inproceedings{ghigliotti_shear-reversal_2015,
	address = {Edinburgh, United Kingdom},
	title = {Shear-reversal in non-{Brownian} suspensions : experiments and simulations},
	shorttitle = {Shear-reversal in non-{Brownian} suspensions},
	url = {https://hal.science/hal-01171413},
	booktitle = {{SoftComp} {Topical} {Workshop}: {Dense} {Suspension} {Flow}},
	author = {Ghigliotti, Giovanni and Lobry, Laurent and Gallier, Stany and Blanc, Frédéric and Lemaire, Elisabeth and Peters, François},
	month = jun,
	year = {2015},
}

@article{diz-pita_predatorprey_2021,
	title = {Predator–{Prey} {Models}: {A} {Review} of {Some} {Recent} {Advances}},
	volume = {9},
	issn = {2227-7390},
	shorttitle = {Predator–{Prey} {Models}},
	url = {https://www.mdpi.com/2227-7390/9/15/1783},
	doi = {10.3390/math9151783},
	language = {en},
	number = {15},
	journal = {Mathematics},
	author = {Diz-Pita, Erika and Otero-Espinar, M. Victoria},
	month = jul,
	year = {2021},
	pages = {1783},
}

@article{li_data-driven_2021,
	title = {A data-driven approach for discovering stochastic dynamical systems with non-{Gaussian} {Lévy} noise},
	volume = {417},
	issn = {0167-2789},
	url = {https://www.sciencedirect.com/science/article/pii/S0167278920308319},
	doi = {10.1016/j.physd.2020.132830},

	journal = {Physica D: Nonlinear Phenomena},
	author = {Li, Yang and Duan, Jinqiao},
	month = mar,
	year = {2021},
	pages = {132830},
}

@article{hamzi_learning_2021,
	title = {Learning dynamical systems from data: {A} simple cross-validation perspective, part {I}: {Parametric} kernel flows},
	volume = {421},
	issn = {0167-2789},
	shorttitle = {Learning dynamical systems from data},
	url = {https://www.sciencedirect.com/science/article/pii/S0167278920308186},
	doi = {10.1016/j.physd.2020.132817},

	journal = {Physica D: Nonlinear Phenomena},
	author = {Hamzi, Boumediene and Owhadi, Houman},
	month = jul,
	year = {2021},
	pages = {132817},
}

@article{massar_equilibrium_2025,
	title = {Equilibrium propagation for learning in {Lagrangian} dynamical systems},
	volume = {112},
	issn = {2470-0045, 2470-0053},
	url = {https://link.aps.org/doi/10.1103/smt9-1t1l},
	doi = {10.1103/smt9-1t1l},
	language = {en},
	number = {3},

	journal = {Physical Review E},
	author = {Massar, Serge},
	month = sep,
	year = {2025},
	pages = {035304},
}

@article{gotoda_detection_2014,
	title = {Detection and control of combustion instability based on the concept of dynamical system theory},
	volume = {89},
	copyright = {http://link.aps.org/licenses/aps-default-license},
	issn = {1539-3755, 1550-2376},
	url = {https://link.aps.org/doi/10.1103/PhysRevE.89.022910},
	doi = {10.1103/PhysRevE.89.022910},
	language = {en},
	number = {2},

	journal = {Physical Review E},
	author = {Gotoda, Hiroshi and Shinoda, Yuta and Kobayashi, Masaki and Okuno, Yuta and Tachibana, Shigeru},
	month = feb,
	year = {2014},
	pages = {022910},
}

@article{huang_generic_2009,
	title = {Generic behavior of master-stability functions in coupled nonlinear dynamical systems},
	volume = {80},
	copyright = {http://link.aps.org/licenses/aps-default-license},
	issn = {1539-3755, 1550-2376},
	url = {https://link.aps.org/doi/10.1103/PhysRevE.80.036204},
	doi = {10.1103/PhysRevE.80.036204},
	language = {en},
	number = {3},

	journal = {Physical Review E},
	author = {Huang, Liang and Chen, Qingfei and Lai, Ying-Cheng and Pecora, Louis M.},
	month = sep,
	year = {2009},
	pages = {036204},
}

@incollection{murray_origin_2018,
	address = {Cambridge},
	series = {Cambridge {Planetary} {Science}},
	title = {The {Origin} of {Planetary} {Ring} {Systems}},
	isbn = {9781107113824},
	url = {https://www.cambridge.org/core/books/planetary-ring-systems/origin-of-planetary-ring-systems/2ED0C5B884A5D9B35E3A366BB176DA96},

	booktitle = {Planetary {Ring} {Systems}: {Properties}, {Structure}, and {Evolution}},
	publisher = {Cambridge University Press},
	author = {Charnoz, S. and Canup, R. M. and Crida, A. and Dones, L.},
	editor = {Murray, Carl D. and Tiscareno, Matthew S.},
	year = {2018},
	doi = {10.1017/9781316286791.018},
	pages = {517--538},
}

@article{canup_origin_2010,
	title = {Origin of {Saturn}’s rings and inner moons by mass removal from a lost {Titan}-sized satellite},
	volume = {468},
	copyright = {2010 Springer Nature Limited},
	issn = {1476-4687},
	url = {https://www.nature.com/articles/nature09661},
	doi = {10.1038/nature09661},
	language = {en},
	number = {7326},
	journal = {Nature},
	author = {Canup, Robin M.},
	month = dec,
	year = {2010},
	pages = {943--946},
}

@article{lumme_formation_1972,
	title = {On the formation of {Saturn}'s rings},
	volume = {15},
	copyright = {http://www.springer.com/tdm},
	issn = {0004-640X, 1572-946X},
	url = {http://link.springer.com/10.1007/BF00649769},
	doi = {10.1007/BF00649769},
	language = {en},
	number = {3},
	journal = {Astrophysics and Space Science},
	author = {Lumme, Kari},
	month = mar,
	year = {1972},
	pages = {404--414},
}

@article{ingersoll_cassini_2020,
	title = {Cassini {Exploration} of the {Planet} {Saturn}: {A} {Comprehensive} {Review}},
	volume = {216},
	issn = {1572-9672},
	shorttitle = {Cassini {Exploration} of the {Planet} {Saturn}},
	url = {https://doi.org/10.1007/s11214-020-00751-1},
	doi = {10.1007/s11214-020-00751-1},
	language = {en},
	number = {8},
	journal = {Space Science Reviews},
	author = {Ingersoll, Andrew P.},
	month = oct,
	year = {2020},
	pages = {122},
}

@article{spilker_cassini-huygens_2019,
	title = {Cassini-{Huygens}’ exploration of the {Saturn} system: 13 years of discovery},
	volume = {364},
	issn = {0036-8075, 1095-9203},
	shorttitle = {Cassini-{Huygens}’ exploration of the {Saturn} system},
	url = {https://www.science.org/doi/10.1126/science.aat3760},
	doi = {10.1126/science.aat3760},
	language = {en},
	number = {6445},

	journal = {Science},
	author = {Spilker, Linda},
	month = jun,
	year = {2019},
	pages = {1046--1051},
}

@article{farmer_understanding_2006,
	title = {Understanding the behavior of {Prometheus} and {Pandora}},
	volume = {180},
	copyright = {https://www.elsevier.com/tdm/userlicense/1.0/},
	issn = {00191035},
	url = {https://linkinghub.elsevier.com/retrieve/pii/S0019103505004008},
	doi = {10.1016/j.icarus.2005.10.005},
	language = {en},
	number = {2},

	journal = {Icarus},
	author = {Farmer, Alison J. and Goldreich, Peter},
	month = feb,
	year = {2006},
	pages = {403--411},
}

@article{poulet_dynamical_2001,
	title = {Dynamical evolution of the {Prometheus}--{Pandora} system},
	volume = {322},
	issn = {0035-8711, 1365-2966},
	url = {https://academic.oup.com/mnras/article/322/2/343/963192},
	doi = {10.1046/j.1365-8711.2001.04128.x},
	language = {en},
	number = {2},

	journal = {Monthly Notices of the Royal Astronomical Society},
	author = {Poulet, F. and Sicardy, B.},
	month = apr,
	year = {2001},
	pages = {343--355},
}

@article{brilliantov_steady_2018,
	title = {Steady oscillations in aggregation-fragmentation processes},
	volume = {98},
	issn = {2470-0045, 2470-0053},
	url = {https://link.aps.org/doi/10.1103/PhysRevE.98.012109},
	doi = {10.1103/PhysRevE.98.012109},
	language = {en},
	number = {1},

	journal = {Physical Review E},
	author = {Brilliantov, N. V. and Otieno, W. and Matveev, S. A. and Smirnov, A. P. and Tyrtyshnikov, E. E. and Krapivsky, P. L.},
	month = jul,
	year = {2018},
	pages = {012109},
}

@article{williams_roche_2003,
	title = {The {Roche} {Limit}},
	volume = {87},
	issn = {1572-9478},
	url = {https://doi.org/10.1023/A:1026137401540},
	doi = {10.1023/A:1026137401540},
	language = {en},
	number = {1},

	journal = {Celestial Mechanics and Dynamical Astronomy},
	author = {Williams, I. P.},
	month = sep,
	year = {2003},
	pages = {13--25},
}

@article{rein_stochastic_2010,
	title = {Stochastic orbital migration of small bodies in {Saturn}’s rings},
	volume = {524},
	copyright = {© ESO, 2010},
	issn = {0004-6361, 1432-0746},
	url = {https://www.aanda.org/articles/aa/abs/2010/16/aa15177-10/aa15177-10.html},
	doi = {10.1051/0004-6361/201015177},
	language = {en},

	journal = {Astronomy \& Astrophysics},
	author = {Rein, H. and Papaloizou, J. C. B.},
	month = dec,
	year = {2010},
	pages = {A22},
}

@article{cuzzi_evolving_2010,
	title = {An {Evolving} {View} of {Saturn}’s {Dynamic} {Rings}},
	volume = {327},
	issn = {0036-8075, 1095-9203},
	url = {https://www.science.org/doi/10.1126/science.1179118},
	doi = {10.1126/science.1179118},
	
	language = {en},
	number = {5972},

	journal = {Science},
	author = {Cuzzi, J. N. and Burns, J. A. and Charnoz, S. and Clark, R. N. and Colwell, J. E. and Dones, L. and Esposito, L. W. and Filacchione, G. and French, R. G. and Hedman, M. M. and Kempf, S. and Marouf, E. A. and Murray, C. D. and Nicholson, P. D. and Porco, C. C. and Schmidt, J. and Showalter, M. R. and Spilker, L. J. and Spitale, J. N. and Srama, R. and Sremčević, M. and Tiscareno, M. S. and Weiss, J.},
	month = mar,
	year = {2010},
	pages = {1470--1475},
}

@article{cuzzi_saturns_2014,
	title = {Saturn’s {F} {Ring} core: {Calm} in the midst of chaos},
	volume = {232},
	issn = {0019-1035},
	shorttitle = {Saturn’s {F} {Ring} core},
	url = {https://www.sciencedirect.com/science/article/pii/S0019103514000098},
	doi = {10.1016/j.icarus.2013.12.027},

	journal = {Icarus},
	author = {Cuzzi, J. N. and Whizin, A. D. and Hogan, R. C. and Dobrovolskis, A. R. and Dones, L. and Showalter, M. R. and Colwell, J. E. and Scargle, J. D.},
	month = apr,
	year = {2014},
	pages = {157--175},
}

@article{tajeddine_what_2017,
	title = {What {Confines} the {Rings} of {Saturn}?},
	volume = {232},
	issn = {0067-0049, 1538-4365},
	url = {https://iopscience.iop.org/article/10.3847/1538-4365/aa8c09},
	doi = {10.3847/1538-4365/aa8c09},
	number = {2},

	journal = {The Astrophysical Journal Supplement Series},
	author = {Tajeddine, Radwan and Nicholson, Philip D. and Longaretti, Pierre-Yves and Moutamid, Maryame El and Burns, Joseph A.},
	month = oct,
	year = {2017},
	pages = {28},
}

@article{blanc_understanding_2025,
	title = {Understanding the {Formation} of {Saturn}’s {Regular} {Moons} in the {Context} of {Giant} {Planet} {Moons} {Formation} {Scenarios}},
	volume = {221},
	issn = {1572-9672},
	url = {https://doi.org/10.1007/s11214-025-01156-8},
	doi = {10.1007/s11214-025-01156-8},
	language = {en},
	number = {3},
	journal = {Space Science Reviews},
	author = {Blanc, Michel and Crida, Aurélien and Shibaike, Yuhito and Charnoz, Sebastien and El Moutamid, Maryame and Estrada, Paul and Mousis, Olivier and Salmon, Julien and Schneeberger, Antoine and Vernazza, Pierre},
	month = mar,
	year = {2025},
	pages = {35},
}

@article{cook_saturns_1973,
	title = {Saturn's rings—{A} survey},
	volume = {18},
	copyright = {https://www.elsevier.com/tdm/userlicense/1.0/},
	issn = {00191035},
	url = {https://linkinghub.elsevier.com/retrieve/pii/0019103573902145},
	doi = {10.1016/0019-1035(73)90214-5},
	language = {en},
	number = {2},

	journal = {Icarus},
	author = {Cook, A.F. and Franklin, F.A. and Palluconi, F.D.},
	month = feb,
	year = {1973},
	pages = {317--337},
}

@article{dibeh_synchronization_2024,
	title = {Synchronization and limit cycles in a simple contagion model with time delays},
	volume = {470},
	issn = {01672789},
	url = {https://linkinghub.elsevier.com/retrieve/pii/S0167278924003671},
	doi = {10.1016/j.physd.2024.134417},
	language = {en},

	journal = {Physica D: Nonlinear Phenomena},
	author = {Dibeh, Ghassan and El Deeb, Omar},
	month = dec,
	year = {2024},
	pages = {134417},
}

@article{esposito_predatorprey_2012,
	title = {A predator–prey model for moon-triggered clumping in {Saturn}’s rings},
	volume = {217},
	copyright = {https://www.elsevier.com/tdm/userlicense/1.0/},
	issn = {00191035},
	url = {https://linkinghub.elsevier.com/retrieve/pii/S0019103511003812},
	doi = {10.1016/j.icarus.2011.09.029},
	language = {en},
	number = {1},

	journal = {Icarus},
	author = {Esposito, Larry W. and Albers, Nicole and Meinke, Bonnie K. and Sremčević, Miodrag and Madhusudhanan, Prasanna and Colwell, Joshua E. and Jerousek, Richard G.},
	month = jan,
	year = {2012},
	pages = {103--114},
}

@article{el_deeb_higher_2024,
	title = {Higher order mass aggregation terms in a nonlinear predator–prey model maintain limit cycle stability in {Saturn}’s {F} ring},
	volume = {468},
	issn = {01672789},
	url = {https://linkinghub.elsevier.com/retrieve/pii/S0167278924002628},
	doi = {10.1016/j.physd.2024.134311},
	language = {en},

	journal = {Physica D: Nonlinear Phenomena},
	author = {El Deeb, Omar},
	month = nov,
	year = {2024},
	pages = {134311},
}

@article{alrebdi2025saturn,
  title={Saturn's F ring is confined by Prometheus and negative diffusion},
  author={Alrebdi, Abdulelah and Esposito, Larry W},
  journal={Icarus},
  volume={441},
  pages={116709},
  year={2025},
  publisher={Elsevier}
}

@article{esposito2025statistics,
  title={Statistics of Saturn's ring occultations: Implications for structure, dynamics, and origins},
  author={Esposito, Larry W and Colwell, Joshua E and Eckert, Stephanie and Green, Melody R and Jerousek, Richard G and Madhusudhanan, Sreenivas},
  journal={Icarus},
  volume={429},
  pages={116386},
  year={2025},
  publisher={Elsevier}
}

@article{mitsi1998dynamics,
  title={Dynamics of nonlinear oscillators under simultaneous internal and external resonances},
  author={Mitsi, S and Natsiavas, S and Tsiafis, I},
  journal={Nonlinear Dynamics},
  volume={16},
  number={1},
  pages={23--39},
  year={1998},
  publisher={Springer}
}

@inproceedings{french2024saturn,
  title={Saturn's F Ring During 13 Years of Cassini ISS Observations},
  author={French, Robert S and Lessard, Madeline and Jankowski, Alyssa and Hedman, Matthew},
  booktitle={56th Annual Meeting of the Division for Planetary Sciences},
  volume={56},
  pages={204--03},
  year={2024}
}

@article{cuzzi2024saturn,
  title={Saturn’s F ring is intermittently shepherded by Prometheus},
  author={Cuzzi, Jeffrey N and Marouf, Essam A and French, Richard G and Murray, Carl D and Cooper, Nicholas J},
  journal={Science Advances},
  volume={10},
  number={19},
  pages={eadl6601},
  year={2024},
  publisher={American Association for the Advancement of Science}
}

\end{document}